\documentclass[twocolumn,amssymb, amsmath,nobibnotes,nofootinbib,showpacs, aps, prb]{revtex4}
\usepackage[dvips]{graphicx}
\begin{document}
\title{Polaron relaxation and hopping conductivity in LaMn$_{1-x}$Fe$_{x}$O$_3$}
\author{A. Karmakar}
\author{S. Majumdar}
\author{S. Giri}
\affiliation{Department of Solid State Physics and Center for Advaced Materials, Indian Association for the Cultivation of Science, Jadavpur, Kolkata 700 032, India}

\begin{abstract}
Dc and ac transport properties as well as electric modulus spectra have been investigated for the samples LaMn$_{1-x}$Fe$_{x}$O$_3$ with compositions 0 $\leq x \leq$ 1.0. The bulk dc resistivity shows a temperature variation consistent with the variable range hopping mechanism at low temperature and Arrhenius mechanism at high temperatures. The ac conductivity has been found to follow a power law behavior at a limited temperature and frequency region  
where Anderson-localization plays a significant role in the transport mechanism for all the compositions. At low temperatures large dc resistivities and dielectric relaxation behavior   for all the compositions are consistent with the polaronic nature of the charge carriers. Scaling of the modulus spectra shows that the charge transport dynamics is independent of temperature for a particular composition but depends strongly on different compositions possibly   due to different charge carrier concentrations and structural properties. 

\end{abstract}


\maketitle

\section{Introduction}

Physical properties of manganites with perovskite structure are governed by the delicate interplay among charge,  lattice, orbital, and spin degrees of freedom which lead to a wide variety of exotic 
effects. \cite{salamon} The most exciting result in hole doped manganites is the discovery of colossal magnetoresistance (CMR), despite the fact that the origin of CMR is still an open question.  \cite{dagotta} Oxygen stoichiometric LaMnO$_3$ is an antiferromagnetic (AFM) insulating parent compound ($T_{\rm N}$ = 140 K) of a family of materials exhibiting CMR. The magnetic structure of LaMnO$_3$ indicates ferromagnetic (FM) Mn-O-Mn superexchange (SE) interactions within the $ab$ plane and AFM Mn-O-Mn SE coupling along the $c$-axis. \cite{wollan,good} Excess oxygen implants 
Mn$^{4+}$ incorporating ferromagnetism through the double exchange (DE) interaction analogous to the scenario of hole doping. \cite{topfer,joy,marko}

Recently, the magnetic and dc transport properties have been investigated in LaMn$_{1-x}$Fe$_{x}$O$_3$. \cite{tong,kde1,zhou} Tong {\it et al}. suggested an unusual DE interaction between Fe$^{3+}$ and Mn$^{3+}$ ions by taking into account the intermediate spin state of Fe$^{3+}$. \cite{tong} Our recent report on the magnetic properties of LaMn$_{1-x}$Fe$_{x}$O$_3$ exhibits different characteristic features depending on the degree of Fe substitution. \cite{kde1} Dominant FM interaction was observed for $x <$ 0.15 whereas a cluster-glass-like state was 
noticed for $x$ = 0.30. \cite{kde2,kde3} In the case of LaMn$_{0.5}$Fe$_{0.5}$O$_3$ 
a strong annealing temperature ($T_{\rm H}$) dependence of magnetization was reported where FM or AFM feature exists as a result of the formation of Mn-rich or Fe-rich clusters whereas  complex glassy magnetic behavior is due to a uniform distribution of Fe and Mn ions depending on $T_{\rm H}$. \cite{kde4,bhame} The compounds with $x >$ 0.5 show a dominant character of AFM SE  interaction. \cite{zhou} Although detailed magnetic properties are available for  LaMn$_{1-x}$Fe$_{x}$O$_3$, the transport mechanism has not been probed explicitly as a result of Fe substitution. Recently, ac and dc transport mechanism has been investigated in low hole doped La$_{1-x}$Sr$_{x}$MnO$_3$, \cite{seeger,pime} La$_{1-x}$Ca$_{x}$MnO$_3$, \cite{cohn} Ca$_{1-x}$Sr$_{x}$MnO$_3$, \cite{cohn} and Pr$_{1-x}$Ca$_{x}$MnO$_3$. \cite{wang,frei,sich,yamada,bisk} Systematic investigation of conductivity mechanism has not been focused on the Mn-site substitution in $R$MnO$_3$ ($R$ = rare earth) compounds with perovskite structure. 

In the present investigation we report on the measurements of the dc resistivity, the complex ac conductivity, and electric modulus spectra at different temperatures for LaMn$_{1-x}$Fe$_{x}$O$_3$ with composition 0 $\leq x \leq$ 1.0. The results indicate that Anderson-localization plays a significant role in the transport mechanism for all the compositions. Dielectric relaxation and conductivity  indicate polaronic transport mechanism in LaMn$_{1-x}$Fe$_{x}$O$_3$.

\section{Experiment}
The polycrystalline samples LaMn$_{1-x}$Fe$_{x}$O$_3$ with $x$ = 0, 0.15, 0.30, 0.50, 0.70, and 1.0 were prepared by a standard sol-gel technique which is described in our earlier report. \cite{kde1} La$_2$O$_3$ (Aldrich, 99.99\%), Mn (Aldrich, 99+\%), and Fe (Alfa Aesar, 99.998\%) were used as starting materials which were dissolved in hydrated nitric acid for preparing metal nitrate solution. Citric acid was then added to get homogeneous mixure of metal citrate precursor which was dried and heated at 550$^\circ$C. The final heat treatments were performed in the form of pellets (1.1 cm in diameter and approximately 1 mm in thickness) at 1000$^\circ$C for 12 hrs in air followed by furnace cooling down to room temperature. The absence of the trace amount of any impurity phase was confirmed by a BRUKER axs powder x-ray diffractometer (Model no. 8D - ADVANCE). X-ray diffraction patterns are shown in Fig. 1. For $x \leq$ 0.50 the x-ray patterns could be indexed by rhombohedral structure ($R\bar{3}c$) whereas orthorhombic structure ($Pbnm$) is observed for $x \geq$ 0.70 which is consistent with the reported results. \cite{zhou} 

\begin{figure}[t]
\vskip 0.0 cm
\centering
\includegraphics[width = 7 cm]{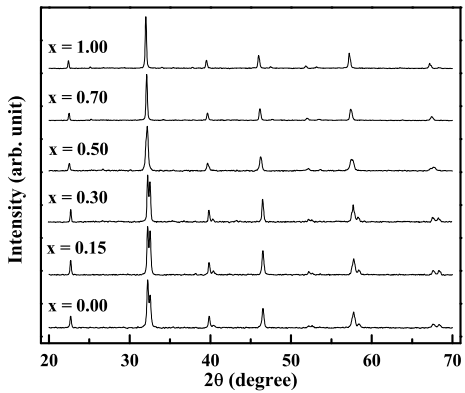}
\caption {X-ray diffraction patterns of LaMn$_{1-x}$Fe$_{x}$O$_3$ with $x$ = 0, 0.15, 0.30, 0.50, 0.70, and 1.0.} 
\label{Fig. 1}
\end{figure}
\begin{figure}[t]
\vskip 0.0 cm
\centering
\includegraphics[width = 8 cm]{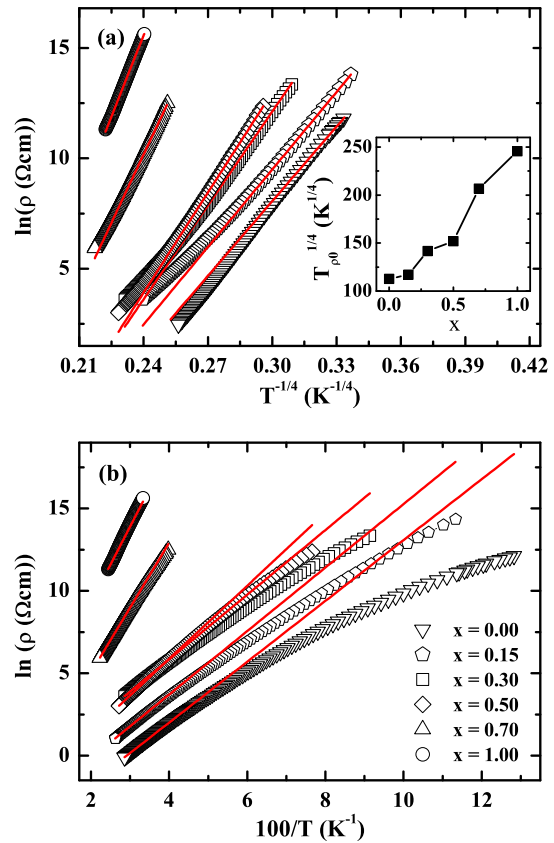}
\caption {(Color online) Variation of the dc resistivity scaled to variable range hoping (VRH) model (top panel) and Arrhenius behavior (bottom panel). The solid straight lines indicate the fits. Inset of the top panel of the figure shows the characteristic temperature, $T_{\rho 0}$ as a function of $x$.}
\label{Fig. 2}
\end{figure}
\par
Capacitance ($C$) and conductance ($G$) measurements were carried out in the frequency range 20 Hz to 2 MHz using an Agilent E4980A LCR meter impedance analyzer which was fitted with a computer for acquiring the online data. The measurement was carried out in the temperature range 20 - 300 K using a low temperature cryo-cooler (Advanced Research Systems, USA) wired with coaxial cables. All samples were circular discs in shape and the electrical contacts were fabricated using air drying silver paint. The silver electrodes along with the sample were cured at 150 $^\circ$C for 4 hours.

\section{Results and discussions}
\subsection{Dc transport}
We have measured dc resistivity ($\rho_{\rm dc}$) with temperature for all the compositions. Semiconducting temperature dependence of resistivity is observed for all the compounds where $\rho_{\rm dc}$ at room temperature increases monotonically with $x$. In case of mixed-valent manganites the conduction of electrons (or holes) is associated with the  polaron formation. If charge carriers are small polarons, $\rho_{\rm dc} (T)$ is expressed as $\rho_{\rm dc} (T)/T$ $\propto$ $\exp[E_{\rm p}/k_{\rm B} T]$; $E_{\rm p}$ being the activation energy. \cite{mott1} We observe that very limited high temperature region could be fitted by the small polaron hoping model. Temperature dependence of $\rho_{\rm dc} (T)$ is also plotted according to Arrhenius formalism, $\rho_{\rm dc} (T)$ $\propto$ $\exp[E_{\rm \rho a}/k_{\rm B} T]$ which is shown in the bottom panel of Fig. 2. Here also a very  limited high temperature region satisfy the Arrhenius behavior. Mott pointed out that at low temperatures the most frequent hopping process would not be to a nearest neighbor. In this temperature regime generally a variable range hopping (VRH)  conductivity sets in. The VRH conduction mechanism can be described by the equation
\begin{equation}
ln(\rho_{\rm dc}/\rho_0) = (T_{\rm \rho 0}/T)^{1/4}
\end{equation}
where $\rho_0$ and $T_{\rm \rho 0}$ are constants. \cite{mott2} Plot of $T_{\rm \rho 0}$ as a function of $x$ is shown in the inset of Fig. 2 where $T_{\rm \rho 0}$ increases with increasing $x$. A change of slope is noticed at $x$ = 0.70 which might be associated with the structural change for $x \geq$ 0.70. The result is also consistent with the estimate of activation energy as a function of $x$ obtained from dc resistivity and dielectric relaxation which are shown in the inset of Fig. 7 where the activation energies from both the calculation exhibits a change of slope for $x \geq$ 0.70. According to VRH mechanism $T_{\rm \rho 0}$ is given by
\begin{equation}
T_{\rm \rho 0} = 24/[\pi k_{\rm B}N(E_{\rm F})\xi^3]
\end{equation}
where $N(E_{\rm F})$ is the localized density of states at the Fermi level and $\xi$ is the decay length of the localized wave function. The activation energy at a particular temperature ($T$) is given by the relation
\begin{equation}
E_{\rho} = 0.25k_{\rm B}T_{\rm \rho 0}^{1/4}T^{3/4}.
\end{equation}
It can be noted that the low temperature data fit well in this scheme which is shown in the top panel of Fig. 2. VRH mechanism normally occurs in the low temperature regime where the temperature is insufficient to excite the charge carriers across the Coulomb gap. In this range conduction takes place normally by hopping of small region ($\approx k_{\rm B}T$) in the vicinity of Fermi energy. Mott's VRH mechanism gives the hopping range of polarons ($R$) 
\begin{equation}
R = \xi^{1/4}/[8\pi k_{\rm B}N(E_{\rm F})T]^{1/4}.
\end{equation}
If $N(E_{\rm F})$ is known $R$ can be determined at a particular temperature. For $x = 0$ the average distance between neighboring Mn ions is $\xi \approx 0.55$ nm. Using Eq. (2) we get $N(E_{\rm F}) = 3.3 \times 10^{18}$ eV$^{-1}$cm$^{-3}$. Thus we get $R \approx 1.57$ nm at $T = 78$ K and $R \approx 1.49$ nm at $T = 98$ K using Eq. (4) where the temperatures  correspond to the range in which resistivity obeys the VRH scheme. We note that the values of $R$ for $x$ = 1 are in the range $\sim$ 2.72 nm ($T$ = 300 K) to $\sim$ 2.63 nm ($T$ = 387 K). The rough  estimate of the range of the distance over which polaron hopping occurs show that $R$ increases considerably with increasing $x$ from $x$ = 0 to 1. 

\begin{figure}[t]
\vskip 0.0 cm
\centering
\includegraphics[width = 8 cm]{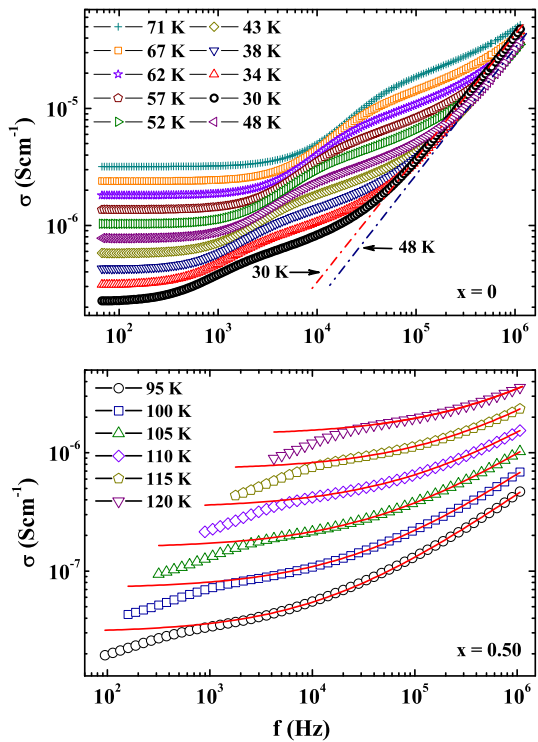}
\caption {(Color online) Frequency dependence of the real part of ac conductivity ($\sigma$)  for $x$ = 0 (top panel) and $x$ = 0.5 (bottom panel). The broken straight line in the top panel indicates the SU behavior at 30 and 48 K. The continuous curves in the bottom panel indicate the fit using UDR behavior.} 
\label{Fig. 3}
\end{figure}

\begin{figure}[t]
\vskip 0.0 cm
\centering
\includegraphics[width = 7 cm]{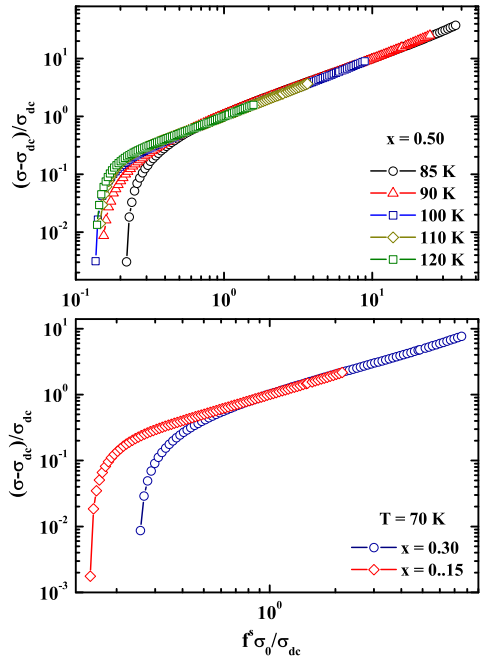}
\caption {(Color online) The plot of $(\sigma - \sigma_{\rm dc})/\sigma_{\rm dc}$ against $f^{\rm s}\sigma_0/\sigma_{\rm dc}$ creating a master curve for $x$ = 0.50 at different temperatures (top panel) and at 70 K for $x$ = 0.15 and 0.30 (bottom panel).} 
\label{Fig. 4}
\end{figure}

\subsection{Ac transport}
 Anderson-localised charge carriers contribute to the conductivity by hopping processes which lead to an increase of complex conductivity with frequency ($f$). This behavior can be described by using a power law with exponent $s <$ 1 which is usually represented as  'universal dielectric response' (UDR). \cite{jons,elliot} By considering a dc conductivity ($\sigma_{\rm dc}$) term the real part of the ac conductivity is described as 
\begin{equation}
\sigma = \sigma_{\rm dc} + \sigma_0f^{\rm s}
\end{equation}
where $\sigma_0$ is a constant. In case of measurement at sufficiently high frequency a cross over from power law, $\sigma \propto$ $f^{\rm s}$ to a linear increase of $f$, $\sigma \propto f$ is frequently observed in a variety of materials which is defined as the 'second universality' (SU). \cite{jons,lee,novick,seeger} The microscopic picture of the origin of SU region at high frequency is still not established.

\begin{figure}[t]
\vskip 0.0 cm
\centering
\includegraphics[width = 8 cm]{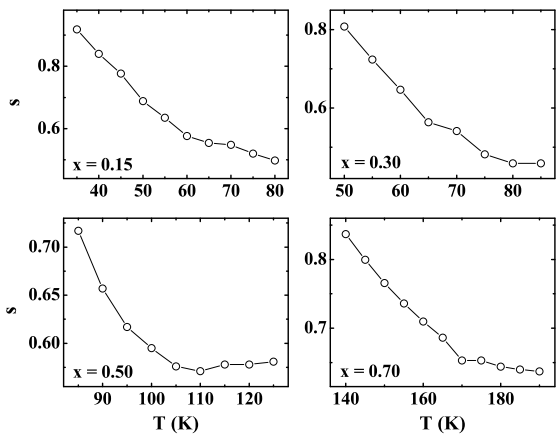}
\caption {Temperature dependence of the exponent, $s$ for $x$ = 0.15, 0.30, 0.50, and 0.70.} 
\label{Fig. 5}
\end{figure}

The frequency dependent complex conductivity has been investigated in LaMn$_{1-x}$Fe$_{x}$O$_3$ for all the compositions. Two of the representations at different temperatures are shown in Fig. 3 for $x$ = 0 and 0.50 in the top and bottom panels of the figure, respectively. At low frequency a step like feature is observed possibly due to the grain boundary effect where a corresponding relaxation peak is observed in the $\epsilon^{\prime\prime}$ spectrum (not shown here). Since our prime concern is understanding the conduction mechanism of the grain interior, we do not resort to an extensive analysis considering the grain boundary contribution. After the step a plateau is noticed corresponding to $\sigma_{\rm dc}$ above which $\sigma$ follows a power law behavior $\propto f^{\rm s}$ in a limited frequency range. With further increasing frequency a cross over to the linear, $\sigma \propto f$ dependence above $\sim$ 10$^5$ Hz is observed at 30 K for $x$ = 0 which is shown in the top panel of Fig. 3. Linearity of the frequency dependence is shown by the broken straight lines in the figure. The linear SU region is shifted toward higher frequencies with increasing temperature and consequently the SU region is shifted out of the frequency window above 48 K for $x$ = 0. The UDR and SU contributions to the intrinsic ac conductivity in the present investigation is consistent with the results for single crystal of LaMnO$_3$ reported by Seeger {\it et al}. \cite{seeger} Since UDR contribution to the ac conductivity is found in a very small region of the frequency, a large error is noticed while analyzing the ac conductivity by the power law for $x$ = 0. The analysis is satisfactory for 0.15 $\leq x \leq$ 0.70 by using Eq. (5) where the low frequency region ascribed to the extrinsic effect is ignored during fitting procedures. One of the examples of the satisfactory fits at different temperatures for $x$ = 0.50 are shown in the bottom panel of Fig. 3 where the low frequency region is not shown explicitly. Eq. (5) can be redefined as 
\begin{equation}
\frac{\sigma - \sigma_{\rm dc}}{\sigma_{\rm dc}} = f^{\rm s} \frac{\sigma_0}{\sigma_{\rm dc}}.
\end{equation}     
According to Eq. (6) the frequency dependence of $\sigma$ at different temperatures can be  scaled to a master curve which confirms the satisfactory fit of the conductivity. \cite{AGhosh} The master curve for $x$ = 0.50 at different temperatures are shown in the top panel of Fig. 4 where the conductivities at high frequency overlap exactly. One can note that at $f^{\rm s} \sigma_0/\sigma_{\rm dc} < \sim$ 0.7 the curves deviate from the linear behavior. This is the region where the UDR behavior cease to exist and is dominated mainly by the grain boundary effect. In the bottom panel of Fig. 4 the conductivities at high frequency also merge to a master curve for $x$ = 0.15 and 0.30 at 70 K where the frequency dependence of both the samples could be fitted satisfactorily at the specified temperature by Eq. (5). Since a very small frequency range could be measured for $x$ = 1.0 satisfying pure UDR region the frequency dependence of the conductivity could not be fitted satisfactorily. We note that the values of $s$ obtained from the fits in the limited temperature region are strongly dependent on temperature for LaMn$_{1-x}$Fe$_{x}$O$_3$ with 0.15 $\leq x \leq$ 0.70 which are shown in Fig. 5. Strong temperature dependence of $s$ has also been reported by Seeger {\it et al}. for low hole doped La$_{1-x}$Sr$_{x}$MnO$_3$. \cite{seeger} In the present investigation temperature dependence of $s$ exhibits similar behavior for different $x$ where $s$ increases approaching toward unity (SU region) with decreasing temperature. The plots in Fig. 5 further indicate that the cross over from UDR to SU region is shifted toward higher temperature with increasing $x$.

\begin{figure}[t]
\vskip 0.0 cm
\centering
\includegraphics[width = 7 cm]{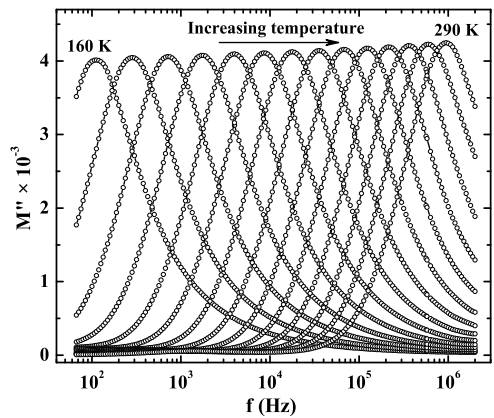}
\caption {Frequency dependence of the imaginary component of the electric modulus at temperatures from 160 K to 290 K at an interval of 10 K for $x$ = 0.7.} 
\label{Fig. 6}
\end{figure}

\subsection{Electric modulus}

Electric modulus corresponds to the relaxation of the electric field in the material when the electric displacement remains constant. Thus the electric modulus represents the real dielectric relaxation process. The complex electric modulus, $M^*$ = $M^{\prime} +  i M^{\prime\prime}$ can be expressed as \cite{macedo}
\begin{equation}
M^*(\omega) 
= M_\infty \left[1-\int^{\infty}_{0}\left(\frac{-d\phi(t)}{dt}\right)\exp\left(-i\omega t\right)dt\right]
\end{equation}
where $M^{\prime}$ and $M^{\prime\prime}$ are the real and imaginary parts of $M^*$. $\omega$ = $2{\pi}f$ is the angular frequency, $M_\infty = 1/\epsilon_\infty$ is the assymptotic value of $M^\prime(\omega)$, and $\phi(t) = \exp\left[-(t/\tau_{\rm M})^\beta\right]$ represents time evolution of the electric field within the material \cite{will} where $\beta$ (0 $\:<\beta\:<$ 1) is the stretched exponent and $\tau_{\rm M}$ is the conductivity relaxation time. The $M^*$ formalism is widely used to study conductivities of materials. The variation of imaginary component, $M^{\prime\prime}$ as a function of frequency at different temperatures provides useful information of the charge transport mechanism such as electrical transport, conductivity relaxation etc. Peaks are  observed in the frequency spectra of $M^{\prime\prime}$ which indicate the existence of conductivity relaxation process. The frequencies, at which the peaks in $M^{\prime\prime}$ spectroscopic plots are observed, follow the relation, $\omega_{\rm max}\tau_{\rm M}$ = 1 where $\omega_{max}$ is the angular frequency corresponding to the peak maximum. It is typically observed that $\tau_{\rm M}$ follows the Arrhenius law given by
\begin{equation}
\tau_{\rm M} = \tau_{\rm M0} \exp(E_{\rm Ma}/k_{\rm B}T)
\end{equation}
where $\tau_{\rm M0}$ is the pre-exponential factor and $E_{\rm Ma}$ the activation energy.

\begin{figure}[t]
\vskip 0.0 cm
\centering
\includegraphics[width = 8 cm]{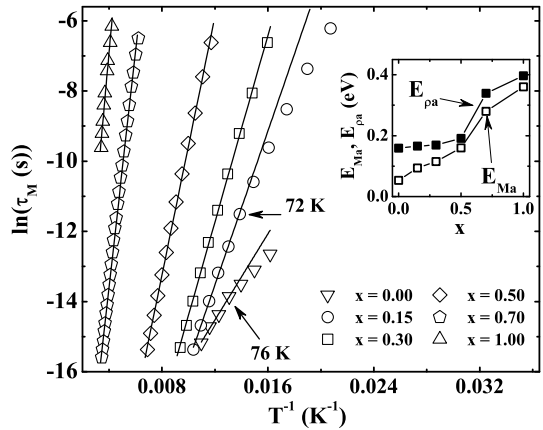}
\caption {The relaxation times ($\tau_{\rm M}$) obtained from the electric modulus spectra are plotted against $T^{-1}$ for all compositions. The solid straight lines are the best fits according to Eq. (8). Insets show the variation of the activation energy, $E_{\rm Ma}$ and $E_{\rm \rho a}$ with $x$ obtained from modulus and dc resistivity, respectively.} 
\label{Fig. 7}
\end{figure}

Figure 6 shows the variation of $M^{\prime\prime}$ with frequency at different temperatures  for a representative composition at $x = 0.7$ where the peaks are shifted toward higher frequencies with increasing temperature. 
We have fitted the variation of the relaxation times ($\tau_{\rm M}$) obtained from the peak frequencies with the inverse of temperature according to the Arrhenius law given by Eq. (8) where the fits are shown by the solid straight lines in Fig. 7. We note that Arrhenius law fits well in the given temperature range for $x \geq$ 0.30 where relaxation peaks are observed in the frequency window until 2 MHz. It deviates below $\sim$ 76 and $\sim$ 72 K for $x$ = 0 and 0.15, respectively (indicated by the arrow in the figure). The deviation from the Arrhenius law at low temperature has also been reported for LaMnO$_3$ \cite{cohn,seeger} and other perovskite material,  Sr$_{0.998}$Ca$_{0.002}$TiO$_3$. \cite{bid} The values of $E_{\rm Ma}$ and $\tau_{\rm M0}$ obtained from the Arrhenius fit are $E_{\rm Ma}$ $\approx$ 53 meV and $\tau_{\rm M0}$ $\approx$  3 $\times 10^{-10}$ s for $x$ = 0 which are in between the reported values 44 meV and $3.3 \times 10^{-8}$ s for polycrystalline, \cite{cohn} and 86 meV and 3 $\times 10^{-13}$ s for single crystalline \cite{seeger} LaMnO$_3$ obtained from dielectric loss spectra. Inset shows the variation of the activation energies as a function of $x$ obtained from the modulus spectra (open symbols) and dc resistivity (filled symbols) satisfying Arrhenius law. The values of activation energies are markedly different at $x$ = 0 where the difference decreases with increasing $x$ and almost close to each other for $x \geq$ 0.50. We note that the slope of  $E_{\rm Ma}$ and $E_{\rm \rho a}$ changes for $x \geq$ 0.70 which might be associated with the structural transition at $x$ = 0.70. The values of $E_{\rm Ma}$ lie in between $53$ meV to $361$ meV while $\tau_{\rm M0}$ are found to be between $\sim$ 3 $\times 10^{-10}$ s and $\sim$ 7 $\times 10^{-13}$ s for $x$ = $0$ to $1$. For comparison $E_{\rm a} = 54$ meV and $\tau_0 = 8.4 \times 10^{-10}$ s were reported for single crystal of CaCu$_3$Ti$_4$O$_{12}$ \cite{homes} and $E_{\rm a} = 313$ meV and $\tau_0 = 8.5 \times 10^{-13}$ s for (Li,Ti)-doped NiO \cite{wu} exhibiting polaronic relaxation process. In the present investigation the values also suggest the typical polaronic relaxation process in LaMn$_{1-x}$Fe$_{x}$O$_3$.

\begin{figure}[t]
\vskip 0.0 cm
\centering
\includegraphics[width = 7 cm]{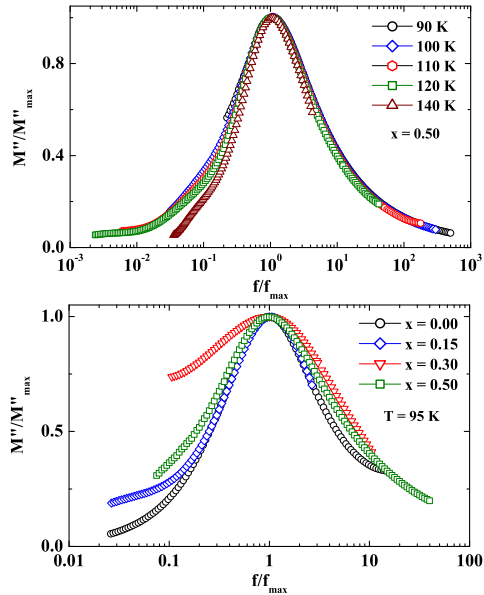}
\caption {(Color online) Scaling of the electric modulus: (top panel) for $x = 0.5$ sample at different temperatures and (bottom panel) for different compositions at 95 K.} 
\label{Fig. 8}
\end{figure}

\begin{figure}[t]
\vskip 0.0 cm
\centering
\includegraphics[width = 9 cm]{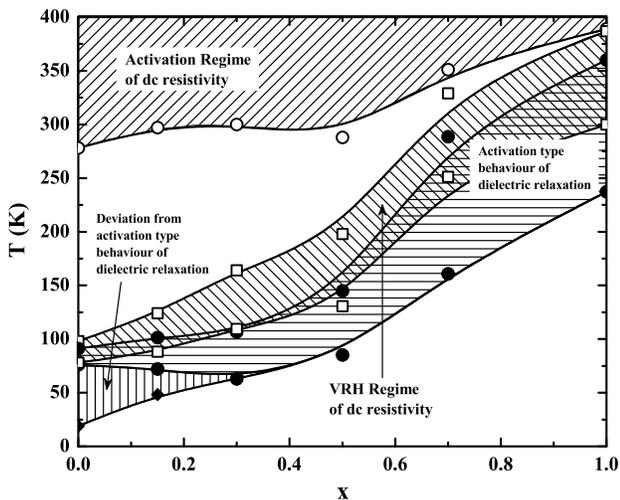}
\caption {Phase diagram from dc resisrivity and dielectric relaxation exhibiting the Arrhenius and VRH regimes in LaMn$_{1-x}$Fe$_x$O$_3$ as a function of $x$. The high temperature regime above the open circles shows the 'activation regime' obtained from dc resistivity. The region covered by the open squares indicated by the inclined hatching shows the 'VRH regime' obtained from dc resistivity. Horizontally hatched region specified by the area within filled circles shows the 'activation regime' obtained from the dielectric relaxation in frequency window 20 Hz to 2 MHz. The small vertically  hatched area indicates the region where the relaxation deviates from the activation law.} 
\label{Fig. 9}
\end{figure}

\par
Scaling of the electric modulus can give further information about the dependence of the relaxation dynamics on the temperature, structure and also on the concentration of the charge carriers. Top panel of Fig. 8 shows the scaling results at different temperatures for the composition at $x = 0.5$ where $M^{\prime\prime}_{\rm max}$ and $\omega_{\rm max}$ are used as the scaling parameters for $M^{\prime\prime}$ and $\omega$, respectively. Clearly all modulus spectra can be seen to completely overlap and are scaled to a single master curve indicating that the relaxation dynamics does not change with temperature for a particular composition at $x$ = 0.50. Bottom panel of Fig. 8 shows the scaling of the electric modulus for different compositions ($x = 0.00, 0.15, 0.30$ and $0.50$) at a fixed temperature, $T$ =  95 K. Data for the other two compositions at $x$ = 0.70 and 1.0 could not be shown in the bottom panel of Fig. 8 because the peaks start originating in these samples at much higher temperatures. It can be noted that the curves do not overlap and are different in nature. The observed deviation at high frequencies occurs due to a change in the dynamical properties of the sample whereas the deviation at the low frequencies is mainly due to extrinsic effects. This failure of matching of the curves at different compositions may be due to a change in the concentration of the charge carries.

\section{Summary and Conclusions}
The results of dc resistivity, ac conductivity, and dielectric relaxation are summarized in a $x-T$ phase diagram given in Fig. 9. 
Dc resistivity of the materials are found to obey Mott's VRH mechanism over a limited  temperature region at low temperature for all the compositions where measurement of resistivity up to the lowest temperature could be performed. The inclined hatched regime covered by the open squares indicates the 'VRH regime' of dc resistivity. The limited high temperature region above the open circles shows the 'activation regime' of dc resistivity. Large gap between the 'activation regime' and 'VRH regime' is observed at $x$ = 0 which decreases with increasing $x$. Both the regimes almost merge at $x$ = 1.0. The 'activation regime' obtained from the dielectric relaxation is spread over a limited temperature region in the frequency window 20 Hz - 2 MHz which is shown by the horizontally hatched area. It is notable that the 'activation regime' from the dielectric relaxation does not overlap with the 'activation regime' obtained from the dc resistivity, rather it overlaps partly with the 'VRH regime'. The results clearly indicate that the dc resistivities obeying the activation behavior do not play any significant role on the dielectric relaxation. We further note that a small vertically hatched area indicated in the phase diagram exhibits the region where relaxation deviates from the  the activation behavior for $x <$ 0.30. 

The most important results of the present investigation is the clear evidence of hopping mechanism in the conductivity behavior. Anderson-localization plays a significant role in the transport mechanism of  LaMn$_{1-x}$Fe$_{x}$O$_3$ for all the compositions. Large dc resistivites for all the compositions are suggested due to the Anderson-localization of polaronic charge carriers. Scaling of the modulus spectrum shows that the charge transport dynamics is independent of temperature but is strongly dependent on the compositions attributed to the different charge carrier concentrations and structural properties. 

\noindent
{\bf Acknowledgement}
S.G. wishes to thank Department of Science and Technology, India (Project No. SR/S2/CMP-46/2003) for the financial support.


\begin{thebibliography} {99}
\bibitem{salamon} M. B. Salamon and M. Jaime, Rev. Mod. Phys. {\bf 73}, 583 (2001).
\bibitem{dagotta} E. Dagotto, Nanoscale Phase Separation and Colossal Magnetoresistance,  (Springer-Verlag, Berlin, November 2002).
\bibitem{wollan} E. O. Wollan and E. O. Koehler, Phys. Rev. {\bf 100}, 545 (1955).
\bibitem{good} J. B. Goodenough, Phys. Rev. {\bf 100}, 564 (1955).
\bibitem{topfer} J. T\"{o}pfer and J. B. Goodenough, J. Solid State Chem. {\bf 130}, 117  
(1997).
\bibitem{joy} P. A. Joy, C. R. Sankar, and S. K. Date, J. Phys.: Condens. Matter. {\bf 14}, 4985  (2002); P. A. Joy, C. R. Sankar, and S. K. Date, J. Phys.: Condens. Matter. {\bf 14},  L663 (2002).
\bibitem{marko} V. Markovich, I. Fita, A. I. Shames, R. Puzniak, E. Rozenberg, Y.  Yuzhelevski, D. Mogilyansky, A. Wisniewski, Y. M. Mukovskii, and G. Gorodetsky, J. Phys.:
Condens. Matter. {\bf 15}, 3985 (2003); I. O. Troyanchuk, V. A. Khomchenko, A. N. Chobot, and H. Szymczak, J. Phys.: Condens. Matter. {\bf 15}, 6005 (2003).
\bibitem{tong} W. Tong, B. Zhang, S. Tan, and Y. Zhang, Phys. Rev. B {\bf 70}, 014422 (2004).
\bibitem{kde1} K. De, R. Ray, R. N. Panda, S. Giri, H. Nakamura, and T. Kohara, J. Magn. Magn. Mater. {\bf 288}, 339 (2005).
\bibitem{zhou} X. -D. Zhou, L. R. Pederson, Q. Cai, J. Yang, B. J. Scarfino, M. Kim, W. B. Yelon, W. J. James, H. U. Anderson, and C. Wang, J. Appl. Phys. {\bf 99}, 08M918 (2006).
\bibitem{kde2} M. Patra, K. De, S. Majumdar, and S. Giri, Eur. Phys. J. B {\bf 58}, 367  (2007).
\bibitem{kde3} K. De, M. Patra, S. majumdar, and S. Giri, J. Phys. D: Appl. Phys. {\bf 40},  7614 (2007). 
\bibitem{kde4} K. De, M. Thakur, A. Manna, and S. Giri, J. Appl. Phys. {\bf 99}, 013908 (2006).
\bibitem{bhame} S. D. Bhame, V. L. Joseph Joly, and P. A. Joy, Phys. Rev. B {\bf 72}, 054426
(2005).
\bibitem{seeger} A. Seeeger, P. Lunkenheimer, J. Hemberger, A. A. Mukhin, V. Yu. Ivanov, A. M. Balbashov, and A. Loidl, J. Phys.: Condens. Matter {\bf 11}, 3273 (1999).
\bibitem{pime} A. Pimenov, Ch. Hartinger, A. Loidl, A. A. Mukhin, V. Yu. Ivanov, and A. M. Balbashov, Phys. Rev. B {\bf 59}, 12 419 (1999).
\bibitem{cohn} J. L. Cohn, M. Peterca, and J. J. Neumeier, Phys. Rev. B {\bf 70}, 214433 (2004).
\bibitem{wang} C. C. Wang and L. W. Zhang, New Journal of Physics {\bf 9}, 210 (2007).
\bibitem{frei} R. S. Freitas, J. F. Mitchell, and P. Schiffer, Phys. Rev. B {\bf 72}, 144429  (2005).
\bibitem{sich} J. Sichelschmidt, M. Paraskevopoulos, M. Brando, R. Wehn, D. Ivannikov, F. Mayr, K. Pucher, J. Hemberger, A. Pimenov, H. -A. Krug von Nidda, P. Lunkenheimer, V. Yu.  Ivanov, A. A. Mukhin, A. M. Balbashov, and A. Loidl, Eur. Phys. J. B {\bf 20}, 7 (2001). 
\bibitem{yamada} S. Yamada, Taka-hisa Arima, and K. Takita, J. Phys. Soc. Jpn. {\bf 68},  3701 (1999).
\bibitem{bisk} N. Biskup, A. de Andr\'{e}s, J. L. Martinez, and C. Perca, Phys. Rev. B {\bf 72} 024115 (2005).   
\bibitem{mott1} N. F. Mott, E. A. Davis, Electronic Processes in Non-Crystalline
Materials, (Oxford, Clarendon, 1971).
\bibitem{mott2} N. F. Mott, J. Non-Cryst. Solids {\bf 1}, 1 (1968).
\bibitem{jons} A. K. Jonscher, Dielectric Relaxation in Solids (London: Chelsea Dielectrics, 1983).
\bibitem{elliot} S. R. Elliott, Adv. Phys. {\bf 36}, 135 (1987); A. R. Long, Adv. Phys. {\bf 31}, 553 (1982).
\bibitem{lee} W. K. Lee, J. F. Liu, and A. S. Nowick, Phys. Rev. Lett. {\bf 67}, 1559 (1991).
\bibitem{novick} A. S. Nowick, A. V. Vaysley, and B. S. Lim, J. Appl. Phys. {\bf 76}, 4429 (1994). 
\bibitem{AGhosh} A. Ghosh and A. Pan, Phys. Rev. Lett. {\bf 84}, 2188 (2000).
\bibitem{macedo} P. B. Macedo, C. T. Moynihan, and R. Bose, Phys. Chem. Glass {\bf 13}, 171 (1972).
\bibitem{will} G. Williams and D. C. Watts, Trans. Faraday Soc. {\bf66}, 80 (1970).
\bibitem{homes} C. C. Homes, T. Vogt, S. M. Shapiro, S. Wakimoto, and A. P. Ramirez, Science {\bf293}, 673 (2001).
\bibitem{wu} J. Wu, C. W. Nan, Y. Lin, and Y. Deng, Phys. Rev. Lett. {\bf89}, 217601 (2002).
\bibitem{bid} O. Bidault, M. Maglione, M. Actis, M. Kchikech, and B. Salce, Phys. Rev. B {\bf 52}, 4191 (1995).
\end{thebibliography}
\end{document}